\documentclass[aip,graphicx]{revtex4-1}
\usepackage[cp1251]{inputenc}
\usepackage[T2A]{fontenc}
\usepackage[english]{babel}
\usepackage{amssymb,latexsym,amsmath,amscd}
\usepackage{graphicx,color,framed}
\begin{document}
%\selectlanguage{russian}
\selectlanguage{english}

\title{A statistical theory of coil-to-globule-to-coil transition of a polymer chain in the mixture of good solvents}

\author{\firstname{Yu.~A.} \surname{Budkov}}
\email[]{urabudkov@rambler.ru}
%\homepage[]{Your web page}
%\thanks{}
%\altaffiliation{}
\affiliation{G.A. Krestov Institute of Solution Chemistry of the Russian Academy of Sciences, Laboratory of NMR spectroscopy and numerical investigations of liquids, Ivanovo, Russia}
\affiliation{National Research University Higher School of Economics, Department of Applied Mathematics, Moscow, Russia}
\affiliation{Lomonosov Moscow State University, Department of Chemistry, Moscow, Russia}

\author{\firstname{ A.~L.} \surname{Kolesnikov}}
%\email[]{bancocker@mail.ru}
%\homepage[]{}
%\thanks{}
%\altaffiliation{}
\affiliation{Institut f\"{u}r Nichtklassische Chemie e.V., Universitat Leipzig, Leipzig, Germany}

\author{\firstname{ M.~G.} \surname{Kiselev}}
%\email[]{}
%\homepage[]{}
%\thanks{}
%\altaffiliation{}
\affiliation{G.A. Krestov Institute of Solution Chemistry of the Russian Academy of Sciences, Laboratory of NMR spectroscopy and numerical investigations of liquids, Ivanovo, Russia}
\affiliation{Lomonosov Moscow State University, Department of Chemistry, Moscow, Russia}
\begin{abstract}
We present an off-lattice statistical model of a single polymer chain in mixed solvent media. Taking into account a polymer conformational entropy, renormalization of solvent composition near the polymer backbone, the universal intermolecular excluded volume and Van-der-Waals interactions within the self-consistent field theory the reentrant coil-to-globule-to-coil transition (co-nonsolvency) has been described in this paper. For convenience we split the system volume in two parts: the volume occupied by the polymer chain and the volume of bulk solution. Considering the equilibrium between two sub-volumes, the polymer solvation free energy as a function of radius of gyration and co-solvent mole fraction within internal polymer volume has been obtained. Minimizing the free energy of solvation with respect to its arguments, we show two qulitatively different regimes of co-nonsolvency. Namely, at sufficiently high temperature a reentrant coil-to-globule-to-coil transition proceeds smoothly. On the contrary, when the temperature drops below a certain threshold value a coil-globule transition occurs in the regime of first-order phase transition, i.e., discontinuous changes of the radius of gyration and the local co-solvent mole fraction near the polymer backbone. We show that, when the collapse of polymer chain takes place, the entropy and enthalpy contributions to the solvation free energy of globule strongly grow. From the first principles of statistical thermodynamics we confirm earlier speculations based on the MD simulations results that the co-nonsolvency is  the essentially enthalpic-entropic effect and caused by enthalpy-entropy compensation. We show that the temperature dependences of solution heat capacity change due to the solvation of polymer chain are in qualitative agreement with the Differential scanning calorimetry data for PNIPAM in aqueous methanol.
\end{abstract}

\maketitle
\section{Introduction}
Co-nonsolvency (insolubility of polymer in a mixture of two good solvents) is one of the most intriguing phenomena in physical chemistry of polymers. Despite the great efforts in both experimental \cite{Shild,Winnik,Zhang,Walter2012,Maekawa,Bischofberger,Hoffman1,Hoffman2,Wang2012} and theoretical \cite{Tanaka2008,Mukherji2013,Kremer2014,Mukherji2015,Kremer2015,Van-der-Vegt2015_2,Freed2015,Sapir2015,Sapir2016} investigations in understanding the co-nonsolvency, the mechanism of this phenomenon remains unclear untill now
\cite{Van-der-Vegt2015_2}.

Basing on the results of the experimental researches devoted to a behavior of the single PNIPAM polymer chain \cite{Shild,Zhang,Winnik} as well as a behavior of PNIPAM hydrogels \cite{Shild,Maekawa}, Tanaka et al. \cite{Tanaka2008} formulated a quasi-chemical model of a single PNIPAM chain in a mixed water-methanol solvent. Authors showed that the co-nonsolvency occurs due to a competition of water and methanol molecules for hydrogen bonding with polymer backbone. Thus, it seemed to be that the co-nonsolvency is caused by the hydrogen bonding between the solvent molecules and monomers. However, in recent papers of Mukherji et al. \cite{Mukherji2013,Kremer2014,Mukherji2015} by using MD computer simulations of Lennard-Jones polymer chain dissolved in two-component Lennard-Jones mixed solvent was shown that co-nonsolvency can take place even in the polymer solutions without hydrogen bonding, but it may be driven by the universal Van-der-Waals and excluded volume interactions only. Authors established that a microscopic parameter which mainly determines an availability of the co-nonsolvency is a difference between energetic parameters of polymer-co-solvent and polymer-solvent attraction, i.e., $\epsilon_{pc}-\epsilon_{ps}$. Moreover, they showed that at the sufficiently large value of this difference (or sufficiently low temperature) the coil-globule transition occurs as a first-order phase transition \cite{Kremer2014,Mukherji2015}. Thus, one can conclude that the co-nonsolvency is a generic physical phenomenon which can be caused by only universal Van-der-Waals and excluded volume  interactions between molecules of solvent species and monomers \cite{Mukherji2015}. The avalability of co-nonsolvency in the mixture N,N dimethylformamide/cyclohexane/polystyrene additionally indicates on the correctness of the latter conclusion \cite{Wolf78}.

Mukherji et al. interpreted results of their MD simulation by using a simple analytical lattice adsorption model \cite{Hill,Prigogine}, taking into account so-called bridging mechanism which implies that one co-solvent molecule can be strongly associated with two monomers. In other words, from authors' point of view the co-nonsolvency usually occurs due to an association of co-solvent molecules with the polymer backbone. It should be noted that within such interpretation 'bridging'-induced coil-globule transition is similar to the electrostatically driven coil-globule transition in polyelectrolyte solutions \cite{Brilliantov1998,Schiessel1998,Cherstvy2010}. Recently, basing on the full atomistic MD simulation of the PNIPAM chain in mixed water-methanol solvent, thorough analysis of entropy and enthalpy contributions to the solvation free energy at the level of linear response approximation for the frozen coil and globule states of polymer chain was provided \cite{Van-der-Vegt2015_2}. Authors showed that in the region of co-nonsolvency, when a collapse of polymer coil takes place the energetics of electrostatic, hydrogen bonding, or bridging-type interactions with the globule is found to play no role. Instead, preferential methanol binding results in a significant increase of the globule configurational entropy, stabilizing methanol-enriched globular structures over wet globular structures in neat water \cite{Van-der-Vegt2015_2}. Thus, there is an ambiguity in the interpretation of the co-nonsolvency microscopic mechanism.

However, the main goal of this paper is not to consider the microscopic mechanism of the co-nonsolvency, but to understand its thermodynamic nature more deeply. To reach our goal, we calculate the solvation free energy of the polymer chain as well as its enthalpic and entropic contributions as the functions of mixed solvent composition in a region of co-nonsolvency. To the best of our knowledge, this problem has not been considered from the first principles of the statistical thermodynamics till now. In order to consider the above-mentioned problem we develop the off-lattice statistical model of the single polymer chain in mixed binary solvent. Taking into account the conformational entropy of polymer chain and renormalization of the co-solvent mole fraction near the polymer backbone, we show that the co-nonsolvency can be successfully described within our self-consistent field theory. We show that, when the collapse of polymer chain takes place, the entropy and enthalpy contributions to the solvation free energy of globule strongly grow, almost compensating each other. Thus we obtain that from thermodynamic point of view the co-nonsolvency is the essentially enthalpic-entropic effect and caused by so-called enthalpy-entropy compensation \cite{Vegt2006}. We also show that at sufficiently high temperature a reentrant coil-to-globule-to-coil transition proceeds smoothly. On the contrary, when the temperature drops below a certain threshold value, the coil-globule transition occurs in the regime of first-order phase transition, i.e., discontinuous changes of the radius of gyration and the local co-solvent mole fraction that confirms earlier results of MD simulations \cite{Mukherji2013,Kremer2014,Mukherji2015}. We calculate the solution heat capacity change due to the solvation of polymer chain as a function of the temperature and show that it is in qualitative agreement with the experimental Differential scanning calorimetry (DSC) data for PNIPAM in aqueous methanol \cite{Wang2012}.

\section{Theory}
We consider an isolated polymer chain with a degree of polymerization $N_{m}$ immersed in a low-molecular weight two-component solvent at a specified number density $\rho$ and temperature $T$ that are located at fluid state region. So the polymer chain in our model is dissolved in a mixture of solvent and co-solvent which are good ones for the polymer chain. Thus, a co-solvent concentration in the bulk solution is $\rho x$, while a solvent concentration is $\rho(1-x)$, where $x$ is a co-solvent mole fraction in the bulk. Like in recent works \cite{Budkov1,Budkov2,Budkov3,Odagiri2015} we assume for convenience that the volume of system consists of two parts: the gyration volume $V_{g}=4\pi  R_{g}^3/3$ ($R_{g}$ is a radius of gyration of the polymer chain) containing predominantly monomers of the polymer chain and the bulk solution. To exclude from the consideration a number density change (that can take place near the polymer chain \cite{Budkov2,Budkov3}) which unimportant for this research, we assume that the entire polymer solution is incompressible, so that the solvent number density in the gyration volume can be determined by the relation $\rho_{1}=\rho-\rho_{m}$, where $\rho_{m}=N_{m}/V_{g}$ is a monomer number density. Moreover, we introduce a local co-solvent mole fraction $x_{1}$ by the relations $\rho_{s}=\rho_{1}(1-x_{1})$ and $\rho_{c}=\rho_{1}x_{1}$, where $\rho_{s}=N_{s}/V_{g}$ and $\rho_{c}=N_{c}/V_{g}$ are the local number densities of the solvent and co-solvent in the gyration volume, respectively. We also assume that the pair interaction potentials for monomer-monomer, monomer-solvent, monomer-co-solvent, solvent-solvent, co-solvent-co-solvent, and solvent-co-solvent have a following form
\begin{equation}
\label{eq:pot}
V_{ij}(\bold{r})=\Biggl\{
\begin{aligned}
-\epsilon_{ij}\left(\frac{\sigma_{ij}}{r}\right)^6, \quad& |\bold{r}|> \sigma_{ij}\,\\
\infty,\quad&|\bold{r}|\leq \sigma_{ij},
\end{aligned}
\end{equation}
where $i,j=m,s,c$; $r=|\bold{r}|$; $\sigma_{ij}$ and $\epsilon_{ij}$ are effective diameters and energetic parameters, respectively.
As well as in our previous work \cite{Budkov3}, we assume that $\sigma_{ij}=(\sigma_{ii}+\sigma_{jj})/2$, whereas each energetic
parameter $\epsilon_{ij}$ is considered as independent. Within the present study we do not introduce the second virial
coefficients as the parameters of interaction \cite{Budkov1,Budkov2,Dzubiella2013}, but as in the work \cite{Budkov3} we construct the total free energy by using different expressions which are straightforwardly related to repulsive and attractive parts of interaction potentials (\ref{eq:pot}). Moreover, in contrast to the works \cite{Budkov1,Dzubiella2013} we take into account the both solvent species explicitly.

A conditional solvation free energy of the polymer chain can be written as:
\begin{equation}
\nonumber
\Delta G_{p}(R_{g},N_{s},N_{c})=\mathcal{F}_{id}(R_{g},N_{s},N_{c})+\mathcal{F}_{ex}(R_{g},N_{s},N_{c})
\end{equation}
\begin{equation}
\label{eq:DeltaG}
+PV_{g}-\mu_{s}N_{s}-\mu_{c} N_{c},
\end{equation}
where $R_{g}$ is the radius of gyration of the polymer chain, $N_{s}$ and $N_{c}$ are molecule numbers of the solvent and co-solvent in the gyration volume, respectively;
$\mathcal{F}_{id}(R_{g},N_{s},N_{c})$ is the ideal free energy of the polymer chain and mixed solvent which can be calculated in the following way
\begin{equation}
\label{eq:Fid}
\mathcal{F}_{id}(R_{g},N_{s},N_{c})=\frac{9}{4}k_{B}T\left(\alpha^{2}+\frac{1}{\alpha^2}\right)\nonumber
\end{equation}
\begin{equation}
+N_{s}k_{B}T\left(\ln{\frac{N_{s}\Lambda_{s}^3}{V_{g}}}-1\right)+N_{c}k_{B}T\left(\ln{\frac{N_{c}\Lambda_{c}^3}{V_{g}}}-1\right),
\end{equation}
where $\alpha=R_{g}/R_{0g}$ is the expansion factor, $R_{0g}^2=N_{m}b^2/6$ is the mean-square radius of gyration of the ideal Gaussian polymer chain, $b$ is the Kuhn length of the segment, $k_{B}$ is the Boltzmann constant, $T$ is the absolute temperature, $\Lambda_{s}$ and $\Lambda_{c}$ are the de Broglie wavelengths of the solvent species. The first term in (\ref{eq:Fid}) is the free energy of the ideal Gaussian polymer chain within the Fixman approximation \cite{Fixman,Grosberg,Birshtein}; $P$ is a pressure in the bulk solution which will be determined below. The excess free energy of polymer solution takes the form
\begin{equation}
\label{eq:ev}
\mathcal{F}_{ex}(R_{g},N_{s},N_{c})=\mathcal{F}_{ev}(R_{g},{N}_{s},N_{c})+\mathcal{F}_{att}(R_{g},{N}_{s},N_{c}),
\end{equation}
where $\mathcal{F}_{ev}$ is a contribution of the repulsive interactions in the gyration volume due to the excluded volume of the monomers and molecules of solvent species which we determine through the Mansoori-Carnahan-Starling-Leland equation of state for the hard-spheres mixture (see Supporting information) \cite{Mansoori}. The use of the latter equation of state instead the virial equation of state \cite{Budkov1,Dzubiella2013} allows us to take into account more precisely the packing effects at the large density of solution $\rho$. The contribution of attractive interactions $\mathcal{F}_{att}$ we determine within the standard mean-field approximation as follows
\begin{equation}
\label{eq:att}
\mathcal{F}_{att}(R_{g},N_{s},N_{c})=-\sum\limits_{i,j}\frac{N_{i}N_{j} a_{ij}}{2V_{g}},
\end{equation}
where the interaction parameters $a_{ij}$ can be determined by the standard rule:
\begin{equation}
a_{ij}=\epsilon_{ij}\int\limits_{|\bold{r}|>\sigma_{ij}}d\bold{r} \left(\frac{\sigma_{ij}}{r}\right)^{6}=v_{ij}\epsilon_{ij},
\end{equation}
where the Van-der-Waals volumes $v_{ij}=4\pi\sigma_{ij}^3/3$ are introduced; $i,j=m,s,c$.

We determine the equilibrium values of the expansion factor $\alpha$ and of the local co-solvent mole fraction $x_{1}$ by the minimization of solvation free energy $\Delta G_{p}(\alpha,x_{1})$ (see Supporting information). It is worth noting that for the practical calculations the local co-solvent mole fraction $x_{1}$ may be related to the excess coordination number of co-solvent $\Delta N_{c}=\left(\rho_{1}x_{1}-\rho x\right)V_{g}$ which is usually used to quantify the stability of given polymer state and describe the preferential binding of co-solvent {\sl vs.} solvent to the polymer surface in real polymer solutions \cite{Smith2008}.

\section{Numerical results}

Turning to the numerical calculations, we introduce the dimensionless parameters: $\tilde{T}=k_{B}T/\epsilon_{ss}$, $\tilde{\rho}=\rho b^3$, $\tilde{P}=Pb^3/\epsilon_{ss}$, $\tilde{\epsilon}_{ij}=\epsilon_{ij}/\epsilon_{ss}$, $\tilde{\sigma}_{ij}=\sigma_{ij}/b$. Following the papers of Mukherji et al. \cite{Kremer2014,Mukherji2015}, we use the effective diameters of species: $\tilde{\sigma}_{ss}=\tilde{\sigma}_{cc}=0.5$, $\tilde{\sigma}_{mm}=1$. The latter choice approximately corresponds to the PNIPAM chain dissolved in the water-methanol mixed solvent. To get the co-nonsolvency regime, is needed to take the polymer-co-solvent energetic parameter larger than other ones. Thus, we choose the following values of the energetic parameters: $\tilde{\epsilon}_{cc}=\tilde{\epsilon}_{sc}=\tilde{\epsilon}_{mm}=\tilde{\epsilon}_{ms}=0.5$, $\tilde{\epsilon}_{mc}=1.5$, so that $\tilde{\epsilon}_{mc}-\tilde{\epsilon}_{ms}=1$. We also assume that the degree of polymerization of the polymer chain $N_{m}=10^{2}$.

Figures 1(a,b) illustrate the dependences of expansion factor $\alpha$ and co-solvent mole fraction $x_{1}$ in the gyration volume on the bulk co-solvent mole fraction $x$ at the different values of temperature $\tilde{T}$ under the fixed pressure $\tilde{P}=0.35$. As is seen, at sufficiently high temperature at increasing co-solvent mole fraction $x$ the reentrant coil-to-globule-to-coil transition proceeds smoothly, while the co-solvent mole fraction $x_{1}$ in the gyration volume monotonically increases. However, when the temperature drops below a certain threshold value, the polymer chain undergoes the coil-globule transition in a regime of first-order phase transition. Namely, when the discontinuous decrease in the expansion factor takes place, the local mole fraction of co-solvent in the gyration volume simultaneously abruptly increases. At further increase in the mole fraction of co-solvent in the bulk solution, the expansion factor and the mole fraction of co-solvent in the gyration volume smoothly increase. It is worth noting that increase in the temperature leads to the shift of threshold co-solvent mole fraction at which the coil-globule transition takes place to higher values. This trend qualitatively agrees with the results of the both experiment and MD computer simulation \cite{Walter2012}. We would also like to stress that an availability of the region where the increase in bulk co-solvent mole fraction has only minor effect on the local solvent/co-solvent composition is a natural consequence of the co-solvent molecules' preferential binding with the polymer backbone. Indeed, in the case of strong attractive interactions between polymer and co-solvent, the co-solvent-induced coil-globule transition accompanied by the significant increase of the co-solvent concentration within the polymer volume \cite{Dzubiella2013,Cherstvy2010,Budkov1}. So when the polymer chain adopts a compact globular conformation, further increase in the bulk co-solvent mole fraction should lead to only minor increase in the local co-solvent concentration due to the excluded volume effects. Moreover, the above-mentioned behavior of the local co-solvent mole fraction is in qualitative agreement with the results of MD simulations of conformation transition of PNIPAM hydrogel as a function of the methanol mole fraction in water/methanol mixtures \cite{Walter2012}.

As already pointed out above, the abrupt increase in the local co-solvent mole fraction additionally confirms the results of works \cite{Dzubiella2013,Cherstvy2010,Budkov1} that in the case of strong polymer-co-solvent attraction the co-solvent molecules must be enriched in the globule. Nevertheless, an implicit accounting for the solvent molecules does not allow us to obtain the reentrant coil-to-globule-to-coil transition, but only co-solvent induced coil-globule transition \cite{Dzubiella2013,Budkov1}. In our opinion, it may be related to the fact that implicit solvent models do not take into account the preferential binding co-solvent {\sl vs} solvent to the polymer chain which, it seems, should play a crucial role in the co-nonosolvency.  Moreover, presence of the abrupt decrease in expansion factor predicted by present theory confirms the earlier MD simulations results \cite{Mukherji2013,Kremer2014,Mukherji2015}. Finally before we pass to the discussion of thermodynamic aspects of the co-nonsolvency, it is instructive to discuss how the difference between the energetic parameters of attraction polymer-co-solvent and polymer-solvent $\epsilon_{mc}-\epsilon_{ms}$ influences on the coil-to-globule-to-coil transition. We obtain (see Fig.2) that increase in the difference $\epsilon_{mc}-\epsilon_{ms}$ leads to more pronounced coil-to-globule-to-coil transition that also confirms the recent MD simulations results \cite{Kremer2014,Mukherji2015}. It should be noted that increase in the degree of polymerization $N_{m}$ also leads to more pronounced the coil-to-globule-to-coil transition.

In order to understand a thermodynamic nature of co-nonsolvency, we discuss a behavior of entropic and enthalpic contributions to the solvation free energy of the polymer chain in the region where the reentrant coil-to-globule-to-coil transition takes place. We shall discuss below the solvation free energy per monomer $\Delta G_{p}/N=\Delta g_{p}=\Delta h_{p}-T\Delta s_{p}$, where $\Delta s_{p}=-\partial{\Delta g_{p}}/\partial{T}$ and $\Delta h_{p}=-T^2\partial{\left(\Delta g_{p}/T\right)}/\partial{T}$ are entropy and enthalpy of solvation per monomer, respectively. On the Figures 3a,b are depicted the dependences of solvation entropy and solvation enthalpy on the co-solvent mole fraction at the fixed temperatures $\tilde{T}=0.38$ (fig. 3a) and $\tilde{T}=0.4$ (fig. 3b) and the pressure $\tilde{P}=0.35$ at a region of co-nonsolvency. As is seen, enthalpy $\Delta h_{p}$ and entropy $-T\Delta s_{p}$ contributions are strongly oscillating functions of the co-solvent mole fraction $x$ within the region of co-nonsolvency in both presented cases. For instance, when the co-solvent mole fraction increases, the solvation enthalpy $\Delta h_{p}$ at first monotonically decreases, attains a minimum, abruptly increases to a maximum, and further monotonically decreases. The entropy contribution $T\Delta s_{p}$ behaves analogously. It should be noted that an abrupt increase (decrease) of the enthalpy (entropy) contribution corresponds to the coil-globule transition, occuring in the regime of first-order phase transition. In contrast to the enthalpy and entropy of solvation, the free energy of solvation in the co-nonsolvency region close to zero. The latter means that entropy and enthalpy contributions almost compensate each other. Thus the co-nonsolvency has to be considered as complex thermodynamic process driven by entropy-enthalpy compensation that confirms the speculations presented in works \cite{Kremer2014,Mukherji2015}. It is interesting to discuss the change of solution heat capacity $\Delta C_{p}=\partial{\Delta{H}_{p}}/\partial{T}$ due to the polymer chain solvation as a function of the temperature at different solvent composition $x$ in the co-nonsolvency region. We obtain (see Fig.4) that each curve $\Delta \tilde{C}_{p}=\Delta C_{p}/k_{B}$ has pronounced minimum which decreases with increase in the bulk co-solvent mole fraction $x$. Such behavior of $\Delta \tilde{C}_{p}$ is in qualitative agreement with the experimental data on the heat flow of PNIPAM in aqueous methanol obtained by Differential scanning calorimetry (DSC) method \cite{Wang2012}.

\section{Conclusion}
Taking into account the effects of conformational entropy, renormalizing the solvent composition near the polymer backbone, and universal intermolecular excluded volume and Van-der-Waals interactions within the self-consistent field theory, we have described the reentrant coil-to-globule-to-coil transition of polymer chain in mixture of the good solvents. We show that, when the collapse of polymer chain takes place, the entropy and enthalpy contributions to the solvation free energy of globule strongly grow almost compensating each other. From the first principles of statistical thermodynamics we confirm earlier speculations based on the MD simulations results that the co-nonsolvency is the essentially enthalpic-entropic effect and caused by enthalpy-entropy compensation.

However, we would like to discuss the limitations of the present self-consistent field theory. First, in this work we use the unrealistic potentials of interaction between the particles of solution. In order to apply this theory to real polymer solutions, one can take the more realistic Lennard-Jones potentials using standard Weeks-Chandler-Andersen procedure \cite{Hansen_MacDonald}. However, in present work we have demonstrated only the principle possibility to describe the co-nonsolvency within self-consistent field theory, retaining its application to the experimental systems for the future researches. Second, our theory in present form is based on the assumption that the entire polymer solution is incompressible. Such assumption may be correct, when the polymer solution is under ambient pressure, whereas the incompressibility approximation must be invalid at the region of extremely high pressures \cite{Budkov2}. Recently was experimentally observed \cite{Hoffman1} and confirmed by full atomistic MD simulations \cite{Kremer2015} that co-nonsolvency of PNIPAM in aqueous methanol can be suppressed by application of sufficiently high pressure (order of 500 $MPa$). To describe this very interesting phenomenon theoretically it is necessary to go beyond the incompressibility approximation that is a subject of the forthcoming publications.

\section{Supporting information}
Here we present some calculation details omitted in the main text.
We start from the conditional solvation free energy of polymer chain in the mixed solvent media
\begin{equation}
\nonumber
\Delta G_{p}(R_{g},N_{s},N_{c})=\mathcal{F}_{id}(R_{g},N_{s},N_{c})+\mathcal{F}_{ex}(R_{g},N_{s},N_{c})+PV_{g}-\mu_{s}N_{s}-\mu_{c} N_{c},
\end{equation}
where $V_{g}=4\pi/3 R_{g}^3$ is the volume of gyration of the polymer chain, $N_{s}$ and $N_{c}$ are molecule numbers of the solvent and co-solvent in the gyration volume, respectively; $\mathcal{F}_{id}(R_{g},N_{s},N_{c})$ is the ideal free energy of the polymer chain and mixed solvent which can be calculated in the following way
\begin{equation}
\label{eq:Fid}
\mathcal{F}_{id}(R_{g},N_{s},N_{c})=\frac{9}{4}k_{B}T\left(\alpha^{2}+\frac{1}{\alpha^2}\right)\nonumber
\end{equation}
\begin{equation}
+N_{s}k_{B}T\left(\ln{\frac{N_{s}\Lambda_{s}^3}{V_{g}}}-1\right)+N_{c}k_{B}T\left(\ln{\frac{N_{c}\Lambda_{c}^3}{V_{g}}}-1\right),
\end{equation}
where $\alpha=R_{g}/R_{0g}$ is the expansion factor, $R_{0g}^2=N_{m}b^2/6$ is the mean-square radius of gyration of the ideal Gaussian polymer chain, $b$ is the Kuhn length of the segment, $k_{B}$ is the Boltzmann constant, $T$ is the absolute temperature, $\Lambda_{s}$ and $\Lambda_{c}$ are the de Broglie wavelengths of the solvent species. The first term in (\ref{eq:Fid}) is the free energy of the ideal Gaussian polymer chain within the Fixman approximation; $P$ is the pressure in the bulk solution which will be determined below. The excess free energy of polymer solution takes the form
\begin{equation}
\label{eq:ev}
\mathcal{F}_{ex}(R_{g},N_{s},N_{c})=\mathcal{F}_{ev}(R_{g},{N}_{s},N_{c})+\mathcal{F}_{att}(R_{g},{N}_{s},N_{c}),
\end{equation}
where $\mathcal{F}_{ev}$ is the contribution of the repulsive interactions in the gyration volume due to the excluded volume of the monomers and molecules of solvent species which we determine through the Mansoori-Carnahan-Starling-Leland equation of state for the hard-spheres mixture (see below). The contribution of attractive interactions $\mathcal{F}_{att}$ we determine within the standard mean-field approximation as:
\begin{equation}
\label{eq:att}
\mathcal{F}_{att}(R_{g},N_{s},N_{c})=-\sum\limits_{i,j}\frac{N_{i}N_{j} a_{ij}}{2V_{g}},
\end{equation}
where the interaction parameters $a_{ij}$ can be determined by the standard rule:
\begin{equation}
a_{ij}=\epsilon_{ij}\int\limits_{|\bold{r}|>\sigma_{ij}}d\bold{r} \left(\frac{\sigma_{ij}}{r}\right)^{6}=v_{ij}\epsilon_{ij},
\end{equation}
where the Van-der-Waals volumes $v_{ij}=4\pi\sigma_{ij}^3/3$ are introduced; $i,j=m,s,c$.

Choosing the local mole fraction of co-solvent $x_{1}$ in the gyration volume and the expansion factor $\alpha$ as the order parameters, one can rewrite the solvation free energy in the following way
\begin{equation}
\label{eq:DeltaG2}
\Delta G_{p}(\alpha,x_{1})=\frac{9}{4}k_{B}T\left(\alpha^{2}+\frac{1}{\alpha^2}\right)\nonumber
\end{equation}
\begin{equation}
+\rho_{1}(\alpha)V_{g}(\alpha)k_{B}T\left(x_{1}\left(\ln\left(\rho_{1}(\alpha)x_{1}\Lambda_{c}^3\right)-1\right)+(1-x_{1})\left(\ln\left(\rho_{1}(\alpha)(1-x_{1})\Lambda_{s}^3\right)-1\right)\right)\nonumber
\end{equation}
\begin{equation}
+V_{g}(\alpha)\left(P(\rho,x,T)+f_{ex}(\rho,x_{1},\rho_{m}(\alpha),T)-\rho_{1}(\alpha)\left(\mu_{s}(\rho,x,T)(1-x_{1})+\mu_{c}(\rho,x,T)x_{1}\right)\right),
\end{equation}
where $\rho_{m}(\alpha)=N_{m}/V_{g}(\alpha)=9\sqrt{6}/(2\pi\sqrt{N_{m}}\alpha^3 b^3)$ is a monomer number density and $f_{ex}(\rho,x_{1},\rho_{m},T)$ is a density of excess free energy which has a form
\begin{equation}
f_{ex}(\rho,x_{1},\rho_{m},T)=\rho k_{B}T A(\rho,x_{1},\rho_{m})\nonumber
\end{equation}
\begin{equation}
-\frac{1}{2}\left(a_{pp}\rho_{m}^2+\rho_{1}^2\left(a_{ss}(1-x_{1})^2+a_{cc}x_{1}^2+2a_{sc}(1-x_{1})x_{1}\right)+2\rho_{m}\rho_{1}\left(a_{ms}(1-x_{1})+a_{mc} x_{1}\right)\right),
\end{equation}
where the following short-hand notations are introduced
\begin{equation}
A(\rho,x_{1},\rho_{m})=-\frac{3}{2}\left(1-y_1(\rho,x_{1},\rho_{m})+y_2(\rho,x_{1},\rho_{m})+y_3(\rho,x_{1},\rho_{m})\right)+\frac{3y_2(\rho,x_{1},\rho_{m})+2y_3(\rho,x_{1},\rho_{m})}{1-\xi(\rho,x_{1},\rho_{m})}\nonumber
\end{equation}
\begin{equation}
+\frac{3\left(1-y_1(\rho,x_{1},\rho_{m})-y_2(\rho,x_{1},\rho_{m})-\frac{y_3(\rho,x_{1},\rho_{m})}{3}\right)}{2(1-\xi(\rho,x_{1},\rho_{m}))^2}+(y_3(\rho,x_{1},\rho_{m})-1)\ln(1-\xi(\rho,x_{1},\rho_{m})),
\end{equation}
\begin{equation}
y_{1}(\rho,x_{1},\rho_{m})=\Delta_{cm}\frac{\sigma_{c}+\sigma_{m}}{\sqrt{\sigma_{m}\sigma_c}}+\Delta_{sm}\frac{\sigma_{s}+\sigma_{m}}{\sqrt{\sigma_{m}\sigma_s}}+\Delta_{sc}\frac{\sigma_{s}+\sigma_{c}}{\sqrt{\sigma_{c}\sigma_s}},~\sigma_{i}=\sigma_{ii},
\end{equation}
\begin{equation}
y_2(\rho,x_{1},\rho_{m})=\frac{1}{\xi}\left(\frac{\xi_c}{\sigma_c}+\frac{\xi_s}{\sigma_s}+\frac{\xi_{m}}{\sigma_{m}}\right)\left(\Delta_{cm}\sqrt{\sigma_c\sigma_m}+\Delta_{sm}\sqrt{\sigma_s\sigma_m}+\Delta_{sc}\sqrt{\sigma_{s}\sigma_{c}}\right),
\end{equation}
\begin{equation}
y_{3}(\rho,x_{1},\rho_{m})=\left(\left(\frac{\xi_{c}}{\xi}\right)^{2/3}\left(\frac{\rho_{1}x_{1}}{\rho}\right)^{1/3}+\left(\frac{\xi_{s}}{\xi}\right)^{2/3}\left(\frac{\rho_{1}(1-x_{1})}{\rho}\right)^{1/3}+\left(\frac{\xi_{m}}{\xi}\right)^{2/3}\left(\frac{\rho_{m}}{\rho}\right)^{1/3}\right)^{3},
\end{equation}
\begin{equation}
\label{eq:Delta}
\Delta_{sm}=\frac{\sqrt{\xi_{s}\xi_{m}}}{\xi}\frac{(\sigma_{s}-\sigma_{m})^2}{\sigma_{s}\sigma_{m}}\frac{\sqrt{\rho_{1}\rho_{m}(1-x_{1})}}{\rho},~\Delta_{cm}=\frac{\sqrt{\xi_{c}\xi_{m}}}{\xi}\frac{(\sigma_{c}-\sigma_{m})^2}{\sigma_{c}\sigma_{m}}\frac{\sqrt{\rho_{1}\rho_{m}x_{1}}}{\rho},
\end{equation}
\begin{equation}
\label{eq:Deltacs}
\Delta_{cs}=\frac{\sqrt{\xi_{c}\xi_{s}}}{\xi}\frac{(\sigma_{c}-\sigma_{s})^2}{\sigma_{c}\sigma_{s}}\frac{\rho_{1}}{\rho}\sqrt{x_{1}(1-x_{1})}
\end{equation}
\begin{equation}
\xi_{s}=\frac{\pi \rho _{1}(1-x_{1})\sigma_{s}^3}{6}, ~ \xi_{c}=\frac{\pi  \rho _{1}x_{1}\sigma_{c}^3}{6}, ~ \xi_{m}=\frac{\pi \rho_{m}\sigma_{m}^3}{6}, ~ \rho_{1}=\rho-\rho_m ,
\end{equation}
\begin{equation}
\xi=\xi(\rho,x_{1},\rho_{m})=\xi_{s}+\xi_{c}+ \xi_{m};
\end{equation}
the local solvent composition $x_{1}$ in the gyration volume is introduced by the following relations
\begin{equation}
\rho_{s}=\frac{N_{s}}{V_{g}}=\rho_{1}(1-x_{1}),~\rho_{s}=\frac{N_{c}}{V_{g}}=\rho_{1}x_{1}.
\end{equation}

The pressure in the bulk solution $P$ in our model is determined by the following equation of state:
\begin{equation}
\label{eq:P}
\frac{P(\rho,x,T)}{\rho k_{B}T}= \frac{1+\xi(\rho,x,0)+\xi^2(\rho,x,0)-3\xi(\rho,x,0)(y_1(\rho,x,0)+y_2(\rho,x,0)\xi(\rho,x,0)+\frac{\xi^2(\rho,x,0)y_3(\rho,x,0)}{3})}{(1-\xi(\rho,x,0))^3}\nonumber
\end{equation}
\begin{equation}
-\frac{\rho}{2k_{B}T}(a_{ss}(1-x)^2+a_{cc}x^2+2a_{sc}x(1-x)),
\end{equation}
where the first term in eq. (\ref{eq:P}) determines a pressure of the two-component hard spheres mixture within the Mansoori-Carnahan-Starling-Leland equation of state; the second term determines the contribution of attractive interactions to the pressure within the mean-field approximation. The chemical potentials of the solvent species can be calculated by the following obvious thermodynamic relations
\begin{equation}
\label{eq:muc}
\mu_{c}(\rho,x,T)=\frac{1}{\rho}\left(P(\rho,x,T)+f(\rho,x,T)+(1-x)\left(\frac{\partial{f(\rho,x,T)}}{\partial{x}}\right)_{\rho,T}\right),
\end{equation}
\begin{equation}
\label{eq:mus}
\mu_{s}(\rho,x,T)=\frac{1}{\rho}\left(P(\rho,x,T)+f(\rho,x,T)-x\left(\frac{\partial{f(\rho,x,T)}}{\partial{x}}\right)_{\rho,T}\right),
\end{equation}
where $f(\rho,x,T)$ is a density of Helmholtz free energy of the bulk solution which can be calculated as
\begin{equation}
f(\rho,x,T)=\rho k_{B}T\left(x\ln\left(\rho\Lambda_{c}^3 x\right)+(1-x)\ln\left(\rho\Lambda_{s}^3 (1-x)\right)\right)+\rho k_{B}T A(\rho,x,0)\nonumber
\end{equation}
\begin{equation}
-\frac{1}{2}\rho^2\left(a_{ss}x^2+a_{cc}(1-x)^2+2a_{cs}x(1-x)\right).
\end{equation}

\bibliographystyle{ugost2008}
\begin{acknowledgments}
This work was supported by grant from the President of the Russian Federation (No MK-2823.2015.3). The part concerning development of theoretical model has been supported by Russian Scientific Foundation (grant No 14-33-00017).
\end{acknowledgments}

\newpage

\begin{figure}
\center{\includegraphics[width=0.6\linewidth]{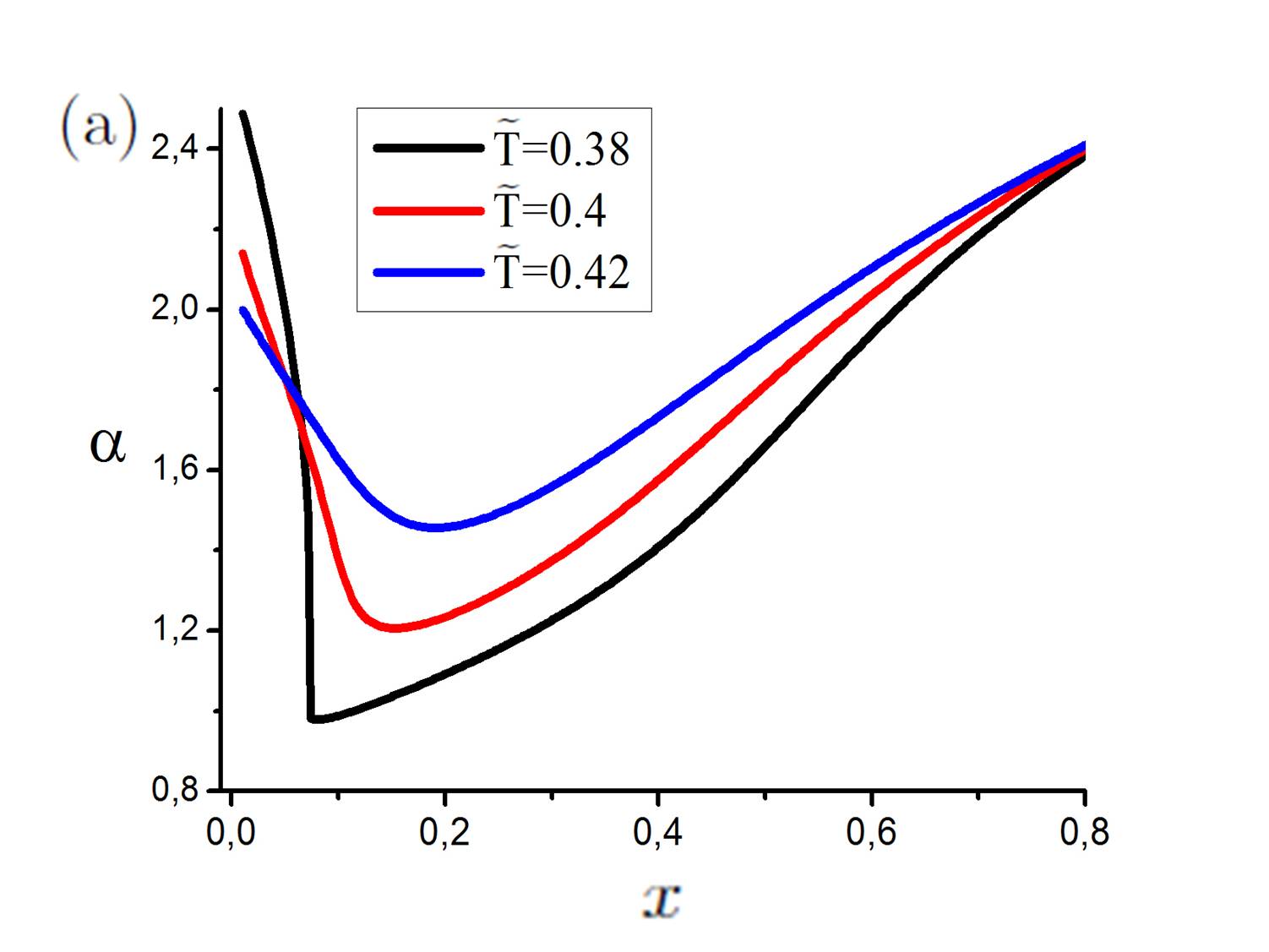}}
\center{\includegraphics[width=0.6\linewidth]{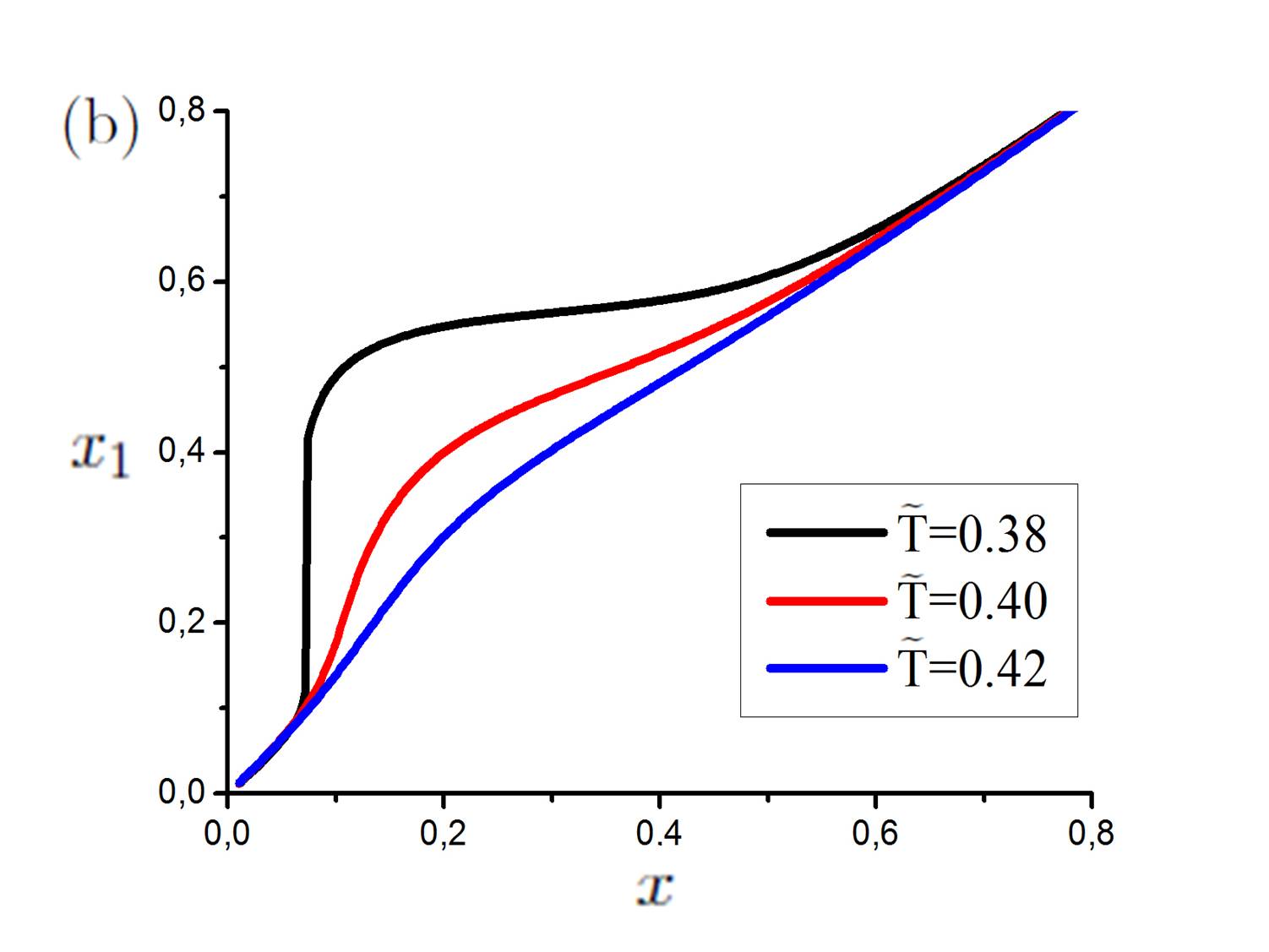}}
\caption{The dependences of expansion factor $\alpha$ (a) and local co-solvent mole fraction $x_{1}$ (b) on the co-solvent mole fraction $x$ in the bulk solution under the fixed pressure $\tilde{P}=0.35$ at the different values of temperature $\tilde{T}$. At sufficiently high temperature at increasing co-solvent mole fraction $x$ the reentrant coil-to-globule-to-coil transition proceeds smoothly, while the co-solvent mole fraction $x_{1}$ in the gyration volume monotonically increases.  However, when the temperature drops below a certain threshold value, the polymer chain undergoes the coil-globule transition in a regime of first-order phase transition. Namely, when the discontinuous decrease of the expansion factor takes place, the local mole fraction of co-solvent in the gyration volume abruptly increases.}
\label{fig.1a,b}
\end{figure}

\begin{figure}
\center{\includegraphics[width=0.6\linewidth]{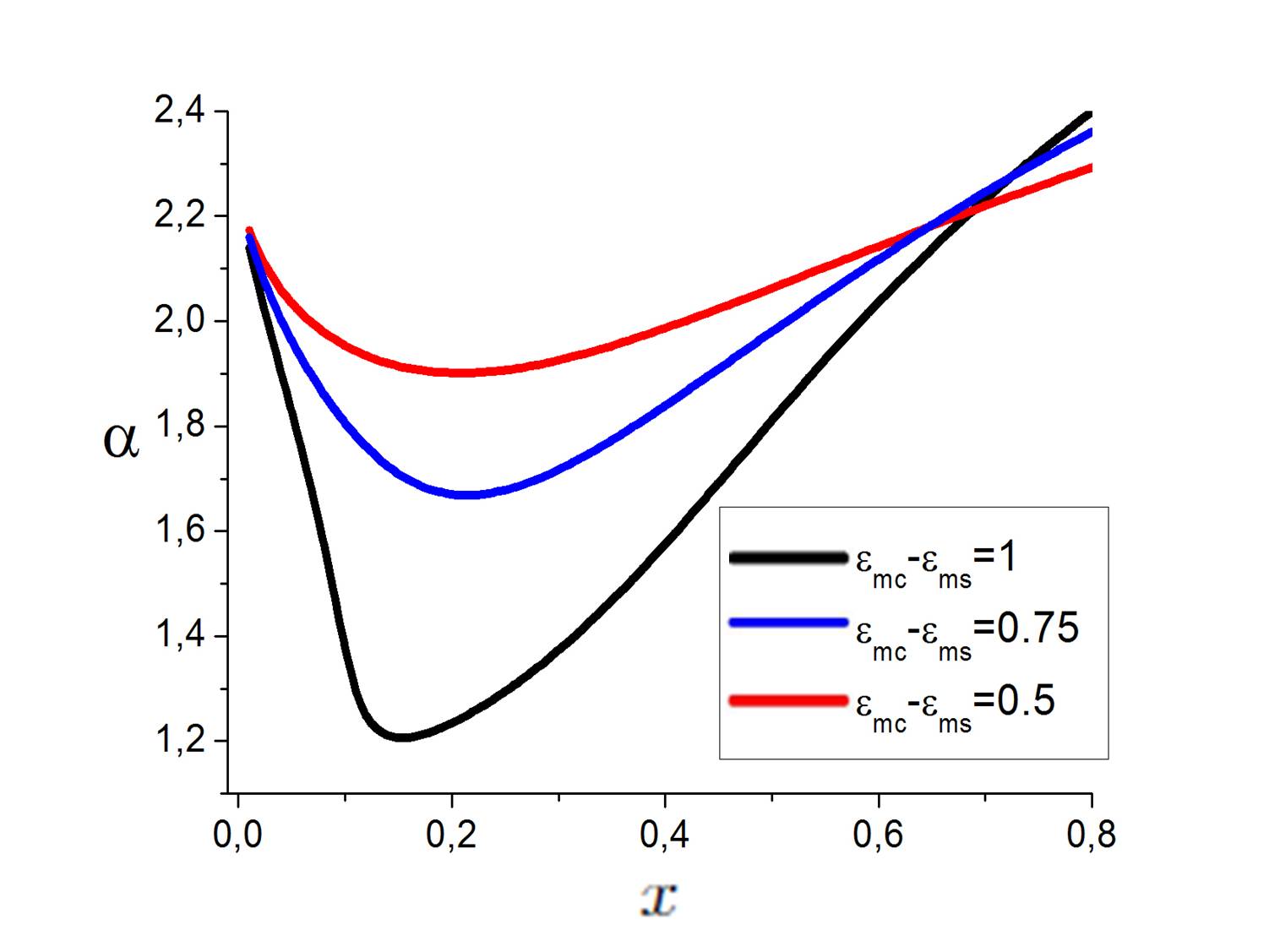}}
\caption{The dependences of expansion factor $\alpha$ on the co-solvent mole fraction $x$ in the bulk solution at the different values $\tilde{\epsilon}_{mc}-\tilde{\epsilon}_{ms}$. Increase in the difference $\tilde{\epsilon}_{mc}-\tilde{\epsilon}_{ms}$ leads to more pronounced coil-to-globule-to-coil transition that also confirms the recent MD simulations results \cite{Kremer2014,Mukherji2015}. The data are shown for $\tilde{T}=0.4$, $\tilde{P}=0.35$, $\tilde{\epsilon}_{ms}=0.5$.}
\label{fig.1a,b}
\end{figure}

\begin{figure}
\center{\includegraphics[width=0.6\linewidth]{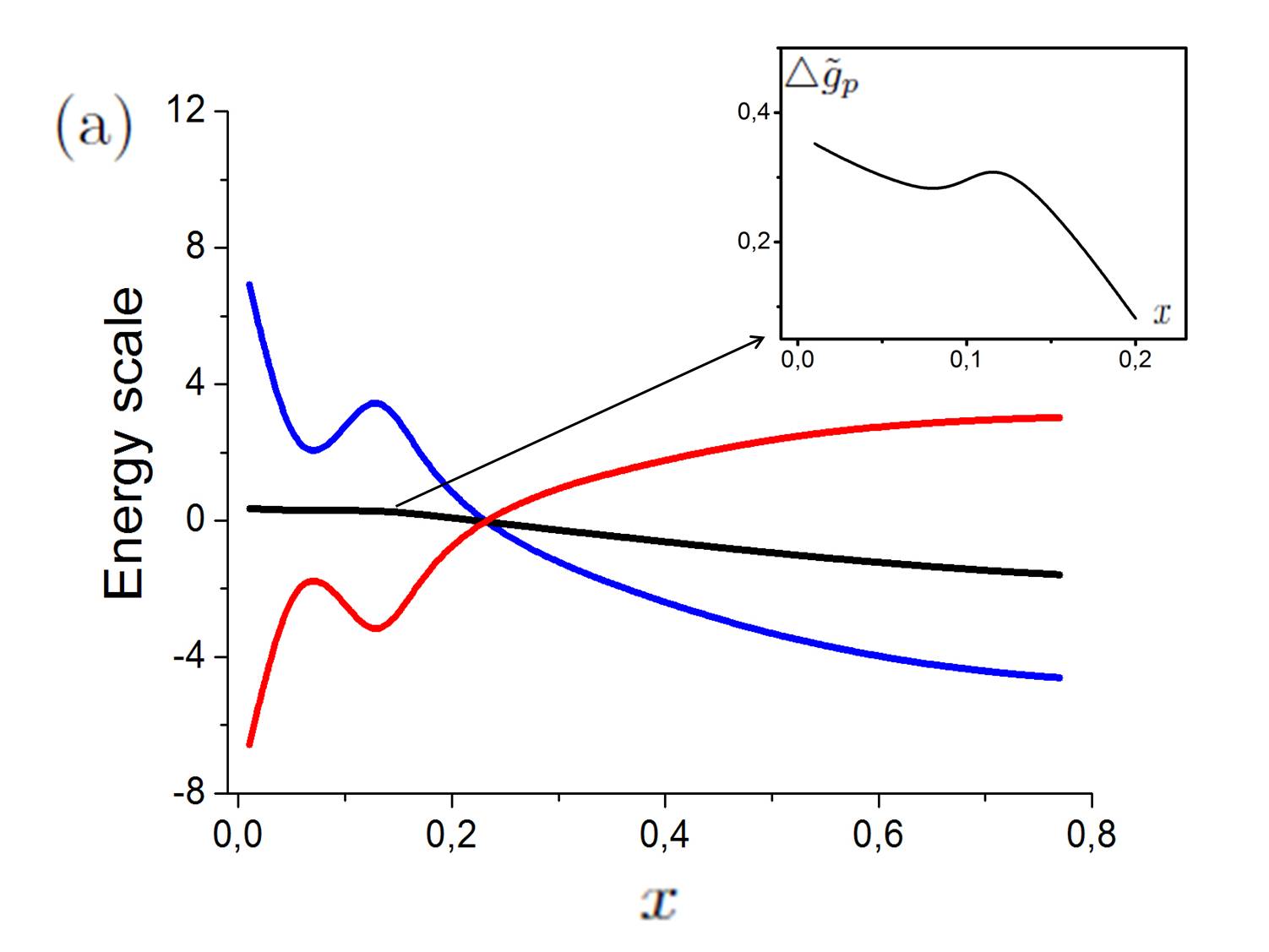}}
\center{\includegraphics[width=0.6\linewidth]{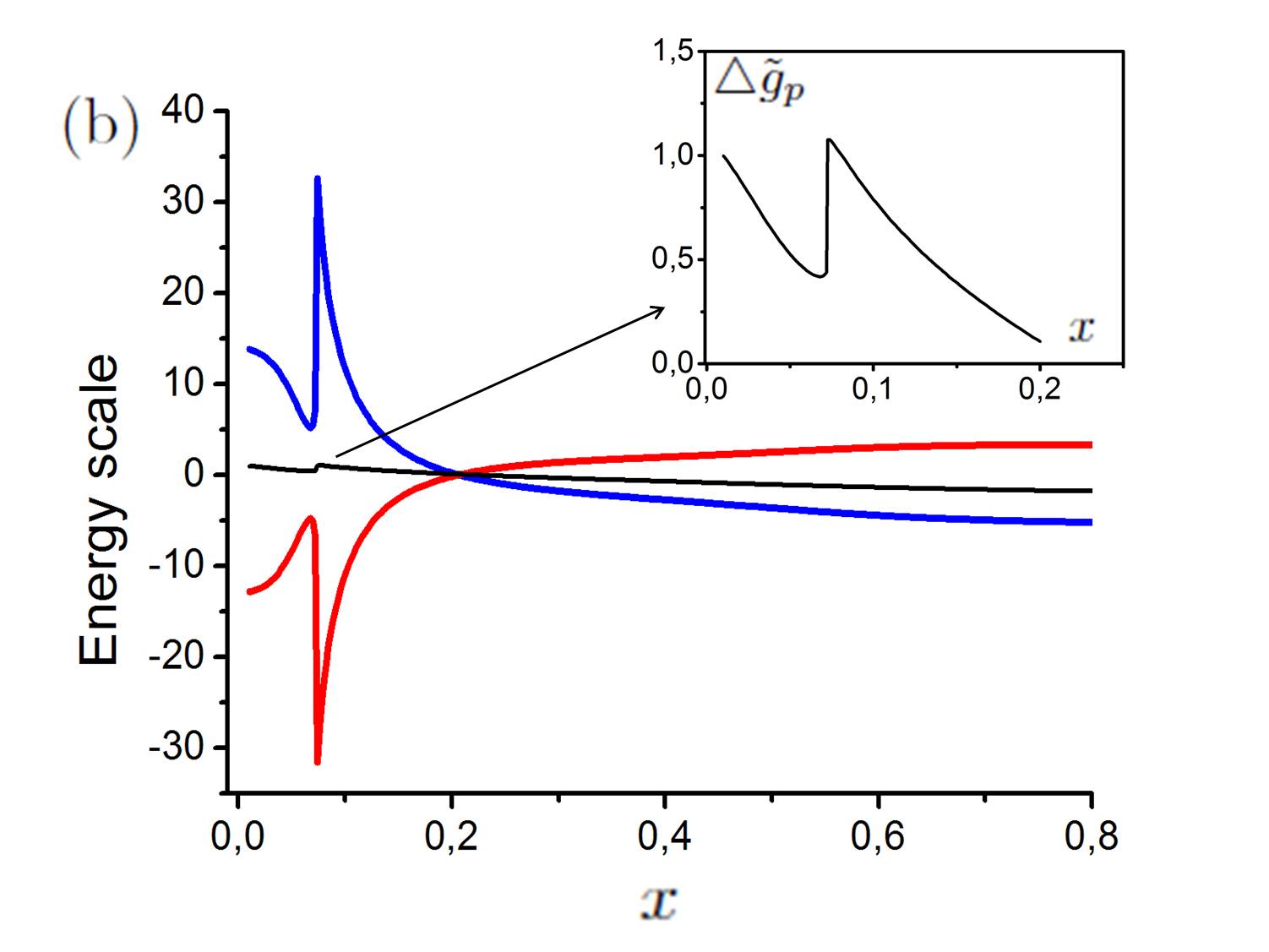}}
\caption{The dependences of the free energy $\Delta \tilde{g}_{p}$ (black lines), the enthalpy $\Delta \tilde{h}_{p}$ (blue lines), and the entropy $-\tilde{T}\Delta \tilde{s}_{p}$ (red lines) of solvation per monomer expressed in units of $\epsilon_{ss}$ on the co-solvent mole fraction $x$ in the bulk solution at the fixed pressure $\tilde{P}=0.35$ and the temperatures (a) $\tilde{T}=0.4$ and (b) $\tilde{T}=0.38$. Enthalpy $\Delta \tilde{h}_{p}$ and entropy $-\tilde{T}\Delta \tilde{s}_{p}$ contributions are strongly oscillating functions of the co-solvent mole fraction $x$ at the region of co-nonsolvency in both cases. The entropy and enthalpy contributions to the solvation free energy almost compensate each other.}
\label{fig.2}
\end{figure}

\begin{figure}
\center{\includegraphics[width=0.6\linewidth]{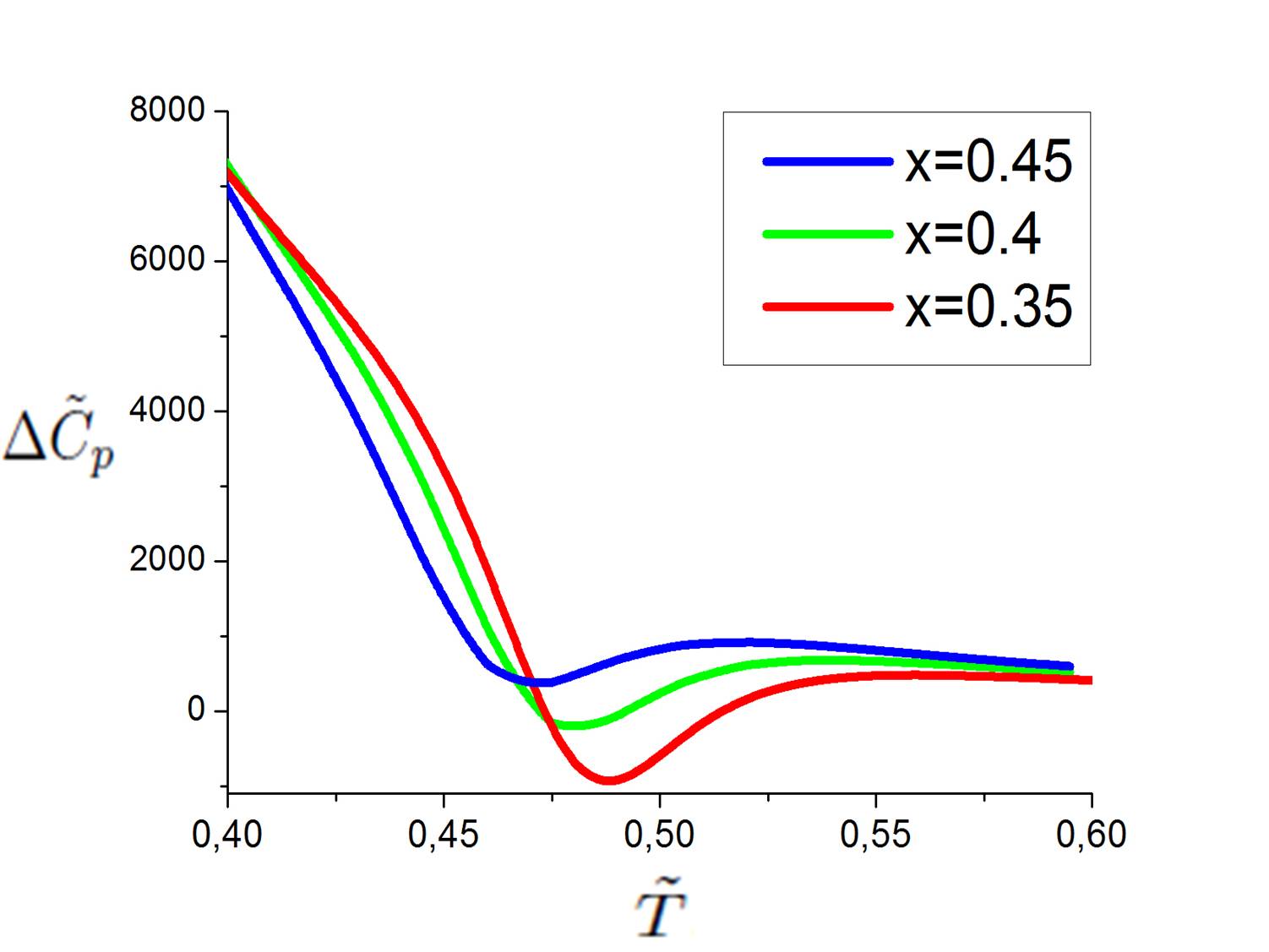}}
\caption{The solution heat capacity change $\Delta \tilde{C}_{p}=\partial{\Delta{\tilde{H}}_{p}}/\partial{\tilde{T}}$ due to the polymer chain solvation as a function of the temperature $\tilde{T}$ at different solvent composition $x$. Each curve $\Delta \tilde{C}_{p}=\Delta C_{p}/k_{B}$ has pronounced minimum which decreases with increase in the bulk co-solvent mole fraction $x$. Such behavior of $\Delta \tilde{C}_{p}$ is in qualitative agreement with the experimental data on the heat flow of PNIPAM in aqueous methanol obtained by Differential scanning calorimetry (DSC) method \cite{Wang2012}. The data are shown for $\tilde{T}=0.4$, $\tilde{P}=0.35$, $\tilde{\epsilon}_{ms}=0.5$, $\tilde{\epsilon}_{mc}=1.5$.}
\label{fig.1a,b}
\end{figure}

\end{document}